\documentclass[twocolumn]{aastex63}

\usepackage[utf8]{inputenc}
\usepackage[english]{babel}
\usepackage{graphicx}
\usepackage{xcolor}
\usepackage{amsmath}

\AtBeginDocument{\usepackage{booktabs}}

\shorttitle{Near $L$-edge photoionization of triply charged iron ions}
\shortauthors{Beerwerth et al.}

\begin{document}

\title{Near $L$-Edge Single and Multiple Photoionization of Triply Charged Iron Ions}

\correspondingauthor{Randolf Beerwerth}
\email{fe3paper@beerwerth.eu}
\correspondingauthor{Stefan Schippers}
\email{stefan.schippers@physik.uni-giessen.de}

\author[0000-0001-5100-4229]{Randolf Beerwerth}
\affiliation{Helmholtz-Institut Jena, Fr\"obelstieg 3, D-07743 Jena, Germany}
\affiliation{Theoretisch-Physikalisches Institut, Friedrich-Schiller-Universit\"at Jena, D-07743 Jena, Germany}

\author[0000-0003-3337-1740]{Ticia Buhr}
\affiliation{I. Physikalisches Institut, Justus-Liebig-Universit\"{a}t Gie{\ss}en, Heinrich-Buff-Ring 16, 35392 Giessen, Germany}

\author[0000-0002-0700-3875]{Alexander Perry-Saßmannshausen}
\affiliation{I. Physikalisches Institut, Justus-Liebig-Universit\"{a}t Gie{\ss}en, Heinrich-Buff-Ring 16, 35392 Giessen, Germany}

\author[0000-0001-9629-9630]{Sebastian O. Stock}
\affiliation{Helmholtz-Institut Jena, Fr\"obelstieg 3, D-07743 Jena, Germany}
\affiliation{Theoretisch-Physikalisches Institut, Friedrich-Schiller-Universit\"at Jena, D-07743 Jena, Germany}

\author[0000-0003-3985-2051]{Sadia Bari}
\affiliation{FS-SCS, DESY, Notkestra{\ss}e 85, 22607 Hamburg, Germany}

\author[0000-0001-8809-1696]{Kristof Holste}
\affiliation{I. Physikalisches Institut, Justus-Liebig-Universit\"{a}t Gie{\ss}en, Heinrich-Buff-Ring 16, 35392 Giessen, Germany}

\author[0000-0002-8805-8690]{A. L. David Kilcoyne}
\affiliation{Advanced Light Source, MS 7-100, Lawrence Berkeley National Laboratory, Berkeley, California 94720, USA}

\author{Simon Reinwardt}
\affiliation{Institut f\"{u}r Experimentalphysik, Universit\"{a}t Hamburg, Luruper Chaussee 149, 22761 Hamburg, Germany}

\author{Sandor Ricz}
\affiliation{Institute for Nuclear Research, Hungarian Academy of Sciences (MTA Atomki), H-4001 Debrecen, Hungary}

\author[0000-0002-1111-6610]{Daniel Wolf Savin}
\affiliation{Columbia Astrophysics Laboratory, Columbia University, 550 West 120th Street, New York, New York 10027, USA}

\author{Kaja Schubert}
\affiliation{FS-SCS, DESY, Notkestra{\ss}e 85, 22607 Hamburg, Germany}

\author[0000-0002-1228-5029]{Michael Martins}
\affiliation{Institut f\"{u}r Experimentalphysik, Universit\"{a}t Hamburg, Luruper Chaussee 149, 22761 Hamburg, Germany}

\author[0000-0002-0030-6929]{Alfred M\"{u}ller}
\affiliation{Institut f\"{u}r Atom- und Molek\"{u}lphysik, Justus-Liebig-Universit\"{a}t Gie{\ss}en, Leihgesterner Weg 217, 35392 Giessen, Germany}

\author[0000-0003-3101-2824]{Stephan Fritzsche}
\affiliation{Helmholtz-Institut Jena, Fr\"obelstieg 3, D-07743 Jena, Germany}
\affiliation{Theoretisch-Physikalisches Institut, Friedrich-Schiller-Universit\"at Jena, D-07743 Jena, Germany}

\author[0000-0002-6166-7138]{Stefan Schippers}
\affiliation{I. Physikalisches Institut, Justus-Liebig-Universit\"{a}t Gie{\ss}en, Heinrich-Buff-Ring 16, 35392 Giessen, Germany}



\begin{abstract}
Relative cross sections for $m$-fold photoionization ($m=1,\ldots,5$) of Fe$^{3+}$ by single photon absorption were measured employing the photon-ion merged-beams setup PIPE at the PETRA\,III synchrotron light source operated at DESY in Hamburg, Germany. The photon energies used spanned the range of $680\text{--}950\,\mathrm{eV}$, covering both the photoexcitation resonances from the $2p$ and $2s$ shells as well as the direct ionization from both shells.
Multiconfiguration Dirac--Hartree--Fock (MCDHF) calculations were performed to simulate the total photoexcitation spectra. 
Good agreement was found with the experimental results. These computations helped to assign several strong resonance features to specific transitions. 
We also carried out Hartree--Fock calculations with relativistic extensions taking into account both photoexcitation and photoionization.
Furthermore, we performed extensive MCDHF calculations of the Auger cascades that result when an electron is removed from the $2p$ and $2s$ shells of Fe$^{3+}$.
Our theoretically predicted charge-state fractions are in good agreement with the experimental results, representing a substantial improvement over previous theoretical calculations.
The main reason for the disagreement with the previous calculations is their lack of inclusion of slow Auger decays of several configurations that can only proceed when accompanied by de-excitation of two electrons.
In such cases, this additional shake-down transition of a (sub-)valence electron is required to gain the necessary energy for the release of the Auger electron.
\end{abstract}





\section{Introduction}

Soft X-ray $L$-shell photoabsorption by $M$-shell iron ions can be important for cosmic objects ranging from photoionized gas in the vicinity of active galactic nuclei (AGNs) to the near neutral gas of the interstellar medium (ISM).
This absorption is largely due to $2p \to 3d$ photoexcitation in Fe$^{0+}$--Fe$^{15+}$, the spectral features of which lie in the $\mathord{\sim}15\text{--}17$~\AA\ bandpass ($\mathord{\sim}730\text{--}830$~eV; \citeauthor{Behar2001} \citeyear{Behar2001}).
To help provide reliable iron $L$-shell photoabsorption data for these astrophysical environments, we have carried out a series of combined experimental and theoretical studies.  
Previously, we presented cross sections for single and multiple photoionization of Fe$^+$ ions in the range of $L$-shell photoexcitation and photoionization \citep{Schippers2017}.
Here we present photoabsorption measurements for Fe$^{3+}$.
Traces of Fe$^{3+}$ may have been detected in AGN spectra \citep[e.g.,][]{Holczer2005}.
In the ISM, Fe$^{3+}$ may also exist in the gas phase \citep{Lee2009}.
But equally important, the Fe in dust grains, when in crystalline structures, may be in the form of Fe$^{3+}$ \citep{Miedema2013}.  
Reliable atomic data for gas-phase Fe$^{3+}$ photoabsorption is needed to distinguish any gas-phase absorption from any solid-matter absorption and for the accurate determination of the iron abundance and its chemical environment.
Benchmarking the relevant ionization cross sections by experimental laboratory studies is a prerequisite for such an analysis, as described in more detail by \citet{Schippers2017}.

Total photoionization cross sections of $L$-shell electrons for iron have been provided by \citet{Reilman1979}.
Theoretical photoionization cross sections for each subshell are tabulated in the works by \citet{Reilman1979}, \citet{Verner1993a}, and \citet{Verner1995}.
Computations of cascade processes that result from inner shell holes were performed and tabulated by \citet{Kaastra1993}, which also includes $L$-shell holes.

Here, we present our measurements of relative cross sections for up to five-fold ionization of Fe$^{3+}$ ions via photoexcitation or photoionization of an $L$-shell electron. Our results provide accurate information on the positions and shapes of the resonances associated with the excitation of a $2p$ electron. These data will help to facilitate a reliable identification of Fe$^{3+}$ photoabsorption features in astrophysical X-ray spectra. Furthermore, we have performed extensive multiconfiguration Dirac--Hartree--Fock (MCDHF) calculations in order to simulate the experimental spectra and to identify the dominant Auger decay channels. 
We also carried out Hartree--Fock calculations with relativistic extensions taking into account both photoexcitation and photoionization.  
Taken together, all these results will be useful for the modeling of the charge balance in astrophysical plasmas.

\begin{figure*}
\includegraphics[width=\textwidth]{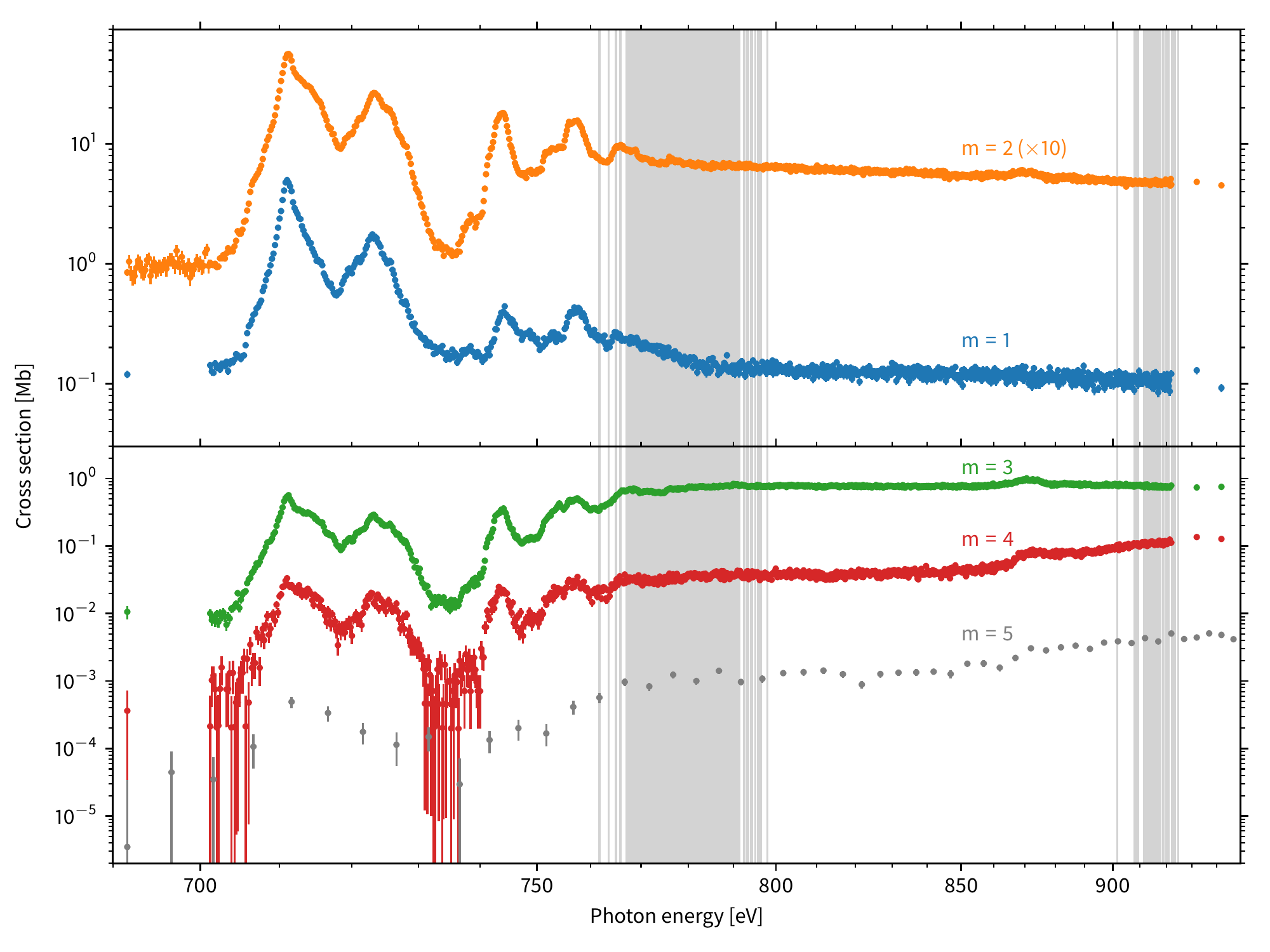}
\caption{\label{fig:cross_sections}Measured partial cross sections, $\sigma_m$, for $m$-fold photoionization of Fe$^{3+}$.
The data are plotted in units of megabarns (Mb), which is $10^{-18}$~cm$^2$.
The partial cross section for $m = 2$ was multiplied by a factor 10 to avoid the large overlap with the $m = 1$ curve. The vertical gray lines show the computed energy level structure of the $2s^2 2p^5 3s^2 3p^6 3d^5$ and $2s 2p^6 3s^2 3p^6 3d^5$ configurations, respectively approximately between 760 and 800~eV and between 900 and 925~eV. The lowest energy levels correspond to the ionization threshold for $2p$ and $2s$ electrons, respectively. For a better view of the low-energy resonance structures, the energy scale has been compressed towards high photon energies according to the formula $E' = \log \left(E - 600 \, \mathrm{e V} \right)$. The absolute cross section scale was obtained by scaling the summed cross section, given by Equation~\eqref{eq:absorption_cross_section}, to the theoretical cross section for photoionization (see text and Figure~\ref{fig:absorption_cross_section}).}
\end{figure*}

\section{Experiment}\label{sec:exp}

The experiment was performed at the PIPE end station \citep{Schippers2014,Mueller2017} of the photon beam line P04 \citep{Viefhaus2013} at the synchrotron light source PETRA III, which is operated by DESY in Hamburg, Germany. At PIPE, the photon-ion merged-beams technique is used to measure photoionization cross sections of ions. 
\citet{Schippers2016} give a recent overview and \citet{Schippers2017} provide a detailed discussion of the experimental method employed here.
Typical Fe$^{3+}$ ion currents in the merged-beam interaction region were $\mathord{\sim} 12$~nA. The nearly monochromatic photon flux, with an energy spread of $\mathord{\sim} 1$~eV, was up to $7.8\times10^{13}$~s$^{-1}$.

Relative cross sections of initial Fe$^{3+}$ ions for the production of Fe$^{q+}$ ions ($4 \leq q\leq 8$) were measured.
As described previously for single and multiple ionization of Fe$^+$ ions \citep{Schippers2017}, these measurements are performed individually for each product charge state $q$ by scanning the photon energy from 680~eV up to 950~eV. 
The results are displayed in Figure~\ref{fig:cross_sections}. 
The measured cross sections span six orders of magnitude.
In our previous work on Fe$^+$, we ruled out contributions to the measured signal due to interactions with more than one photon or ionizing collisions off of the residual gas in the apparatus \citep{Schippers2017}.
There, it was estimated that such events can be safely disregarded. 
Since the present data were obtained under very similar experimental conditions, we attribute the measured cross sections in Figure~\ref{fig:cross_sections} to only processes that involve an initial excitation or ionization of Fe$^{3+}$ by a single photon.

%

\begin{figure}
\includegraphics[width=\columnwidth]{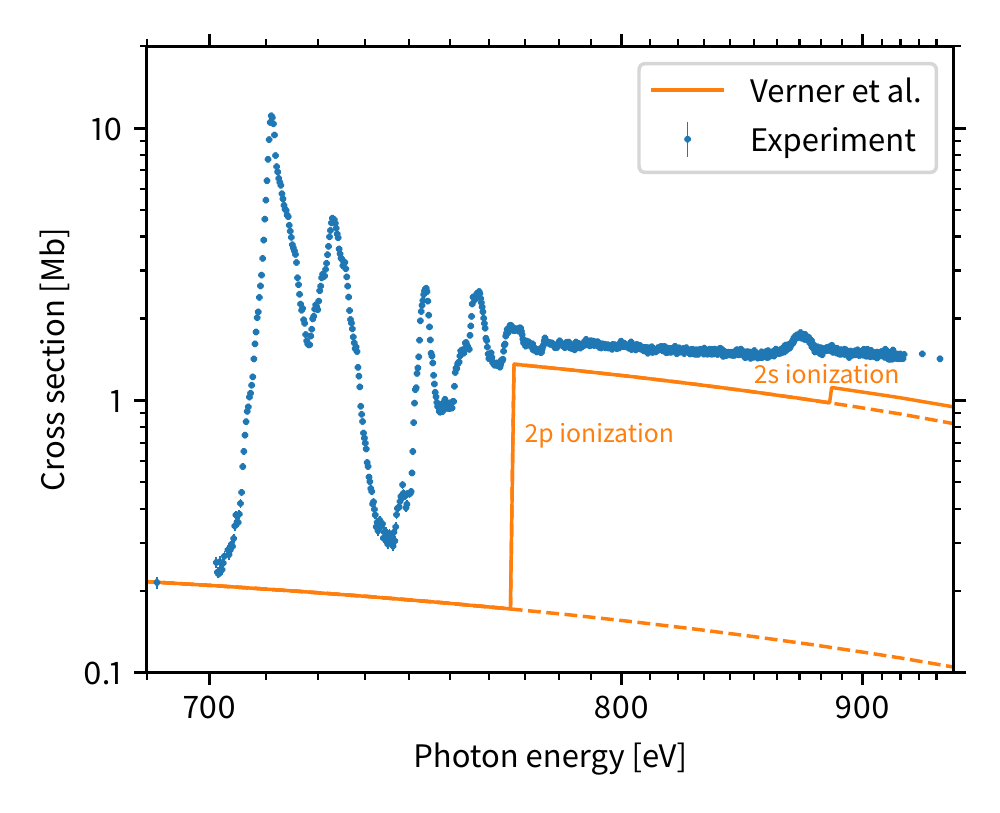}
\caption{\label{fig:absorption_cross_section} Experimental total photoabsorption cross section given by Equation~\eqref{eq:absorption_cross_section} for Fe$^{3+}$ (blue circles) and the theoretical cross section for single-photon single ionization of Fe$^{3+}$ from \citet{Verner1993a} (orange line). 
The steps at $767$~eV and $885$~eV correspond to the thresholds for direction photoionization of a $2p$ and $2s$ electron, respectively.
The dashed lines are the continuation of $M$-shell and $(M+2p)$-shell photoionization.
As in Figure~\ref{fig:cross_sections}, the energy scale is compressed for high energies to enhance the visibility of the low-energy resonance structures.}
\end{figure}

In principle, the PIPE setup enables measuring photoionization cross sections on an absolute scale. This requires scanning the spatial profiles of the ion beam and the photon beam, from which the geometrical beam overlap factor can be obtained. Unfortunately, such measurements could not be carried out because of a technical problem that could not be solved within the allocated beamtime. Therefore, we multiplied all relative partial cross sections by a common factor such that the cross section sum,

\begin{equation}\label{eq:absorption_cross_section}
 \sigma_\Sigma = \sum_{m = 1}^{5} \sigma_m \,,
\end{equation}
matches the theoretical photoionization cross section of \citet{Verner1993a} at 692~eV (Figure~\ref{fig:absorption_cross_section}). At these energies the cross section is dominated by photoionization of the $M$-shell. 
The rationale for this procedure is that we found excellent agreement between experiment and theory in this energy range in our previous work on photoionization of Fe$^+$ where \emph{absolute} cross sections were measured with a $\pm15\%$ total uncertainty at a 90\% confidence limit \citep{Schippers2017}. 
This suggests that there is a similar uncertainty for the absolute cross section scale in the present case, after normalization to the theoretical cross section of \citet{Verner1993a} as described above.
It should be noted that, to a very good approximation, the sum in Equation~\eqref{eq:absorption_cross_section} represents the total photoabsorption cross section, as all the dominant product channels have been measured. 
The unmeasured  Fe$^{3+}$ product channel, which represents photon scattering, is expected to be insignificant because the fluorescence yield from inner shell hole states is generally negligible for light elements like iron \citep{McGuire1972}.
In this case, our computations confirm the fluorescence yield to be about 1\%.

For the determination of the photon energy scale, the same calibration was used as for our Fe$^+$ measurements \citep{Schippers2017}, taking into account the differences in the Doppler shift between the faster Fe$^{3+}$ ions and the slower Fe$^+$ ions. The remaining uncertainty of the experimental photon-energy scale is $\pm$0.2~eV.

The ground level of Fe$^{3+}$ is the $3d^5\;^6S_{5/2}$ level. 
In addition there are 36 excited $3d^5$ levels that can be populated in the hot plasma of the ECR source.
For all these excited levels, the flight time from the ion source to the photon-ion interaction region is much shorter than the radiative lifetime of the levels \citep{Nahar2006,FroeseFischer2008}.
Consequently, the Fe$^{3+}$ ion beam consisted of an unknown mixture of ground-level and excited-level ions. 
This has to be taken into account when comparing the theoretical calculations with the experimental results, as is discussed in more detail below.
Higher-excited even-parity configurations are expected to play a negligible role as their excitation energies are larger than 15~eV.
Therefore, their populations are expected to be insignificant for the ion temperatures inferred below for our ion beam.

\section{Theory} \label{sec:theo}

\begin{figure}
\begin{center}
\includegraphics[width=\columnwidth]{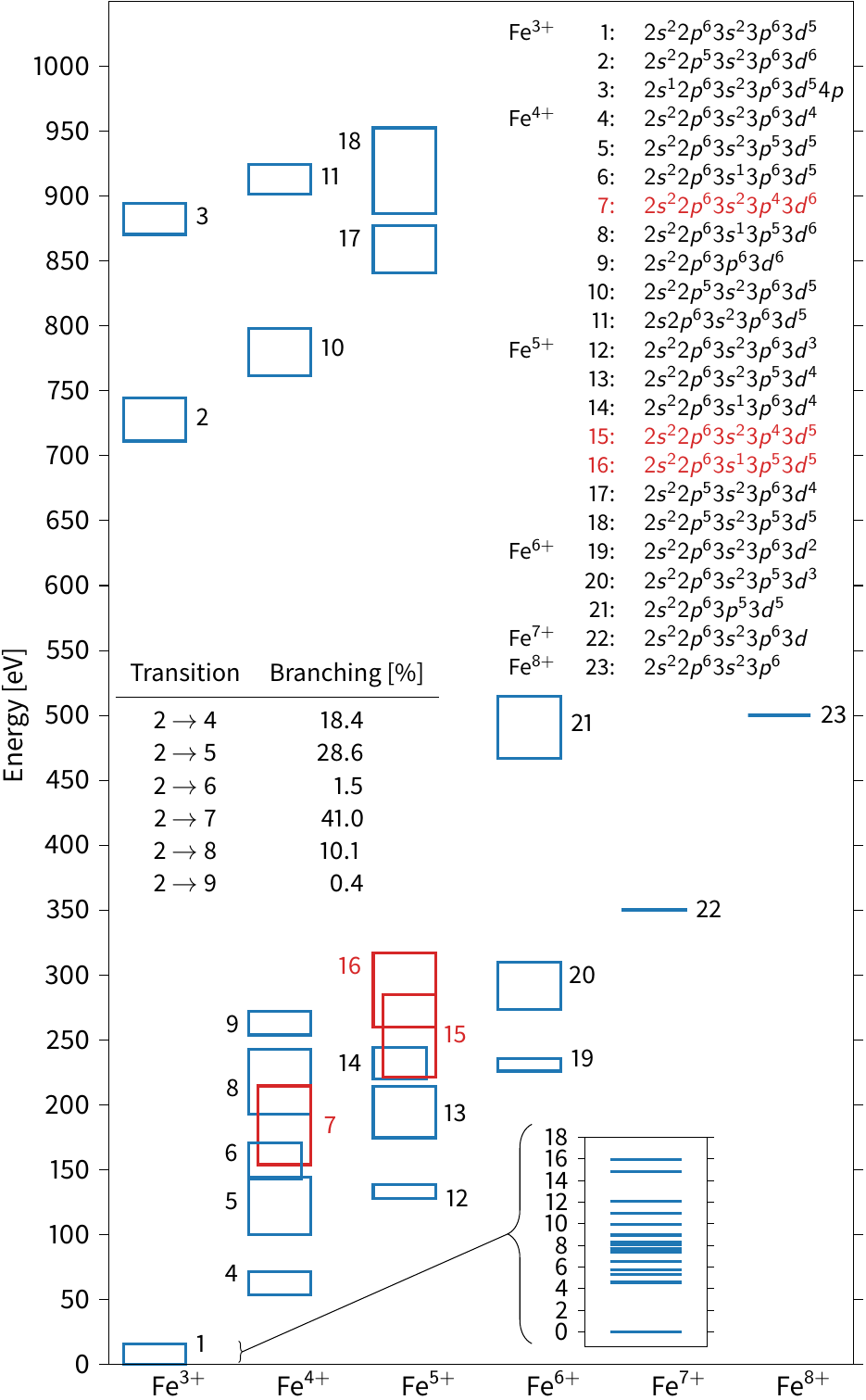}
\end{center}
\caption{\label{fig:level_Diagram}Energy configuration diagram of the hole-state configurations that can be accessed by single-photon excitation or ionization of Fe$^{3+}$. 
Some core-hole configurations that can be accessed with the current photon energies are listed. 
All configurations marked in red can, at least partially, decay via three-electron Auger processes. See Sec.~\ref{sec:cascades} for details.
The inset in the lower right corner shows the computed energy levels of the $2s^2 2p^6 3s^2 3p^6 3d^5$ ground configuration.
The table in the middle gives the branching ratios of the first Auger decay of the $2s^2 2p^5 3s^2 3p^6 3d^6$ configuration formed by photoexcitation.}
\end{figure}

\subsection{MCDHF Calculations} \label{sec:mcdf}

In order to understand and interpret the measured resonance structures, we have performed MCDHF calculations \citep{Grant2007} to model the photoexcitation cross sections. 
The background due to direct photoionization was neglected in these models since the fine-structure resolved absolute photoionization cross sections pose major challenges.
In addition, independent extensive MCDHF computations were performed to model all the de-excitation pathways due to Auger cascade processes of the $2s^{-1}$ and $2p^{-1}$ vacancies created by either photoexcitation or direct photoionization processes.
For all MCDHF computations, we utilized the \textsc{Grasp2k} program package \citep{Joensson2007a, Joensson2013} to generate approximate wave functions, which we describe below. 
The \textsc{Ratip} code was employed to compute all needed transition rates and relative photoionization cross sections \citep{Fritzsche2001, Fritzsche2012a}.

The computed level structure of the $2s^2 2p^6 3s^2 3p^6 3d^5$ ground configuration of Fe$^{3+}$ can be seen in the inset of Figure~\ref{fig:level_Diagram}. The computed gross structure largely reproduces the experimentally derived energy levels (not shown) reported by \citet{Kramida2015}. The most notable observation that can be made here is that the $3d^5\;^6S_{5/2}$ ground level is well separated from the more highly excited metastables. However, we have used here the single configuration approximation without additional corrections for electron correlation effects. As a result, deviations from the measured level energies can be seen. 
For example, the total energy spread of the ground configuration is computed as 16~eV, which is too large by about 2.5~eV \citep{Kramida2015}. Additionally, our computations do not correctly reproduce the level order in some multiplets, due to the limited basis sets used. For example, the first excited ${}^4G$ multiplet has four fine-structure levels ranging from $J = 5/2$ to $J = 11/2$, where the latter is lowest in energy and $J = 7/2$ is highest in energy, separated by about 7.5~meV \citep{Kramida2015}. This order is reversed in our computations, such that $J = 5/2$ comes out lowest and $J = 11/2$ highest. As the fine-structure splitting of 0.01~eV is very small compared to the photon energy spread of 1~eV, an incorrect level order within a multiplet does not affect the computed spectra to any significant extent. Furthermore, we note that our single-configuration computations reproduce reasonably well the lifetimes calculated  by \citet{FroeseFischer2008}.

The photoexcitation cross section due to resonant $2p \rightarrow nd$ photoexcitation was computed based on wave functions for the $3d^5$ ground configuration and the excited $2s^2 2p^5 3s^2 3p^6 \left( 3d^6 + 3d^5 4d + 3d^5 5d  \right)$ configurations, taking limited configuration interaction (CI) into account. 
The contribution of $2p \rightarrow 4s$ photoexcitations into the $2s^2 2p^5 3s^2 3p^6 3d^5 4s$ configuration was found to be negligible and hence has been neglected in the subsequent MCDHF computations.

Inner-shell hole states produced by photoexcitation or photoionization will predominantly decay by Auger processes.
In the most common two-electron Auger process, one electron fills the inner-shell vacancy and the second electron is released into the continuum producing an ion in the next-higher charge state.
A fraction of the Auger decays can result in a so-called shake-up or shake-down transition, where the Auger process is accompanied by an additional excitation or de-excitation, respectively, of a third bound electron, hereafter denoted as a three-electron Auger process.
If instead two electrons are simultaneously ejected into the continuum, the process is called direct double Auger decay.

To model all the de-excitation pathways by sequential Auger decays after, for example, resonant $2p$ photoexcitation of Fe$^{3+}$ (forming configuration 2 in Figure~\ref{fig:level_Diagram}), we include all electronic configurations that arise from two-electron Auger decay processes emerging from the core-hole excited $2s^2 2p^5 3s^2 3p^6 3d^6$ configuration. 
All energetically allowed configurations that emerge in this way are shown in Figure~\ref{fig:level_Diagram}. We note that more configurations might naively be expected to be accessible but cannot be populated by subsequent Auger emissions due to energy conservation in each step. 
Therefore, when direct double-Auger processes as well as shake-up transitions are neglected, $2p \to 3d$ photoexcited ions can only produce ions up to Fe$^{6+}$.
This limitation is due to energy conservation, as the populated levels with the highest energy in the cascade pathways belong to the $3s^{-2}$ configuration in Fe$^{4+}$, labeled 9 in Figure~\ref{fig:level_Diagram}. 
Only photoexcited $2s$ vacancies lie high enough in energy so that their decay can produce ions in the Fe$^{7+}$ charge state in this Auger model.
The Fe$^{8+}$ charge state is not significantly populated for any of the photon energies considered here.
This is also prevented since $3s^{-2}$ vacancies in Fe$^{6+}$ (configuration 21 in Figure~\ref{fig:level_Diagram}) are the highest populated configuration after the decay of a $2p$ vacancy in Fe$^{5+}$ (configurations 17 and 18 in Figure~\ref{fig:level_Diagram}).
Even though a decay to Fe$^{8+}$ is energetically possibly, this fraction is calculated to be around a millionth of a percent and hence orders of magnitude too low to be significant.

Auger cascades resulting from direct photoionization forming Fe$^{4+}$ are modeled in a very similar manner as for resonantly excited Fe$^{3+}$. The Auger cascades that emerge from $2s 2p^6 3s^2 3p^6 3d^5$ and $2s^2 2p^5 3s^2 3p^6 3d^5$ holes are modeled independently. In addition to direct $2s$ and $2p$ photoionization, direct photoionization of an $M$-shell electron and subsequent Auger processes have also been considered. 
As can be seen in Figure~\ref{fig:level_Diagram}, all $2s^2 2p^6 3s 3p^6 3d^5$ holes in Fe$^{4+}$ (configuration 6 in Figure~\ref{fig:level_Diagram}) emit one Auger electron to form Fe$^{5+}$.
However, within the $2s^2 2p^6 3s^2 3p^5 3d^5$ configuration (configuration 5), only the higher-lying levels can undergo an Auger decay to Fe$^{5+}$, while the low-lying levels radiatively relax into the ground configuration of Fe$^{4+}$.

The total non-radiative decay widths of the 180 fine-structure levels of the $2s^2 2p^5 3s^2 3p^6 3d^6$ configuration vary from 370~meV to about 550~meV.
This is expected to be slightly overestimated due to the non-orthogonality of the underlying orbital basis sets for the initial and final wave function expansions.
Within the theoretical accuracy, the total non-radiative decay widths of these $2p$-hole levels created by photoexcitation are similar to the widths of $2p$ vacancies created by direct photoionization, which also vary from 370~meV to about 550~meV.
The $2s$-hole levels can decay by an Auger process where the $2s$ hole is filled by a $2p$ electron, a so-called Coster--Kronig process.
This process is much faster than a typical Auger process.
Hence, as expected from the Heisenberg uncertainty principle, the associated widths of 3.3--3.7~eV are much larger than those of the $2p$-hole levels.
These widths were only computed for Fe$^{4+}$ $2s$-hole levels resulting from direct $2s$ ionization. 
Since the decay widths of $2p$-hole levels in Fe$^{3+}$ and Fe$^{4+}$ are almost identical, it is assumed that this also holds for $2s$ holes. 
Therefore, we assume that the decay widths of $2s$ photoexcited Fe$^{3+}$ levels are within the same range of $2s$ holes in Fe$^{4+}$, formed by direct photoionization of a $2s$ electron in Fe$^{3+}$.

The cascade model that results from the above considerations gives rise to several thousand fine-structure levels for the intermediate charge states, and hence millions of Auger transitions between those levels. In order to keep the calculations of the Auger transition rates tractable, it was necessary to constrain the size of the Auger matrices. Therefore, all wave functions were computed in the single-configuration approximation. 
This approach, detailed in \citet{Buth2018}, neglects effects due to configuration interactions that become crucial for the description of shake-processes as discussed by \citet{Andersson2015} and \citet{Schippers2016a}. 

As an additional simplification to make the calculations more readily tractable, one might consider averaging the transition rates between fine-structure levels of the configurations by assuming a statistical population to obtain an average transition rate between configurations as described by \citet{Buth2018}. However, such an approach yields results that are very similar to the previous computations by \citet{Kaastra1993}, which do not reproduce the experimental findings very well. Therefore, we built the full decay tree between fine-structure levels based on the transition rates computed in the single-configuration approximation, while still neglecting radiative losses as they are much slower than Auger processes. 
Using this approach, we are able to account for the highly non-statistical population of the fine-structure levels of the initial hole configuration due to the photoexcitation or photoionization of Fe$^{3+}$.

\subsection{HFR Calculations}  \label{sec:hfr}

Additional calculations have been performed on a CI level utilizing Hartree--Fock wavefunctions with relativistic extensions (HFR) using the Cowan code \citep{Cowan1981}.
These calculations account for both photoexcitation and photoionization.
CI is included in the initial and the $2p$ photoexcited or photoionized levels. All possible $LS$-levels are taken into account. The lifetimes, i.e., the line widths of the core hole resonances, are calculated from the Auger decay rates to various final Fe$^{4+}$ levels.

For the initial levels the $3d^3 4s^2 + 3d^4 4s + 3d^{5}$ configurations are taken into account, with identical $2s^2 2p^6 3s^2 3p^6$ core configurations. 
Cross sections are calculated for the $2p$ core excitation from initial level configurations into $2s^2 2p^5 3s^2 3p^6 3d^4 4s^2 + 2s^2 2p^5 3s^2 3p^6 3d^5 4s + 2s^2 2p^5 3s^2 3p^6 3d^6$. 
Excitations into Rydberg-like $nd$ ($n \ge 4$) orbitals are not taken into account.

As in the MCDHF calculations for the core excited levels, we calculate the Auger transition rates taking into account the decay into the intermediate Fe$^{4+}$ configurations $3p^4 3d^{6-k} 4s^k \epsilon(s,d)$ and $3p^6 3d^{4-k} 4s^k \epsilon(s,d)$ for outgoing $s$ or $d$ waves and  $3p^5 3d^{5-k} 4s^k  \epsilon(p,f)$ for $p$ and $f$ waves with $k=0, 1, 2$.
Here, the $2s^2 2p^6 3s^2$ core is common for all configurations and $\epsilon$ signifies a free electron. 
Auger decay channels forming a $3s^{-2}$ hole are omitted, due to their low transition rates as confirmed by the computed branching ratios shown in the inset table in Figure~\ref{fig:level_Diagram}.
The calculated lifetime from the Auger transition rates of the core excited levels results in typical line widths in the range of 200--300~meV.

\section{Results and discussion} \label{sec:res}

\begin{deluxetable*}{llllllll}
\tablecaption{\label{tab:Xsec} Measured partial cross sections, $\sigma_m$, for $m$-fold photoionization of Fe$^{3+}$ ions (Figure~\ref{fig:cross_sections}); resulting summed cross section, $\sigma_\Sigma$, given by Equation~\eqref{eq:absorption_cross_section} (Figure~\ref{fig:absorption_cross_section}); and mean product charge-state, $\overline{q}$, given by Equation~\eqref{eq:qmean} (Figure~\ref{fig:total_ion_yield}b). 
The numbers in parentheses in this table provide the one-sigma statistical experimental uncertainties (see also the text below Equation~\eqref{eq:absorption_cross_section}, for a discussion of the systematic uncertainty of the cross section scale).
The systematic uncertainty of the energy scale is $\pm0.2$~eV.}
\tablehead{
    \colhead{Energy~(eV)} &
    \colhead{$\sigma_{1}$~(Mb)}&
    \colhead{$\sigma_{2}$~(Mb)}&
    \colhead{$\sigma_{3}$~(Mb)}&
    \colhead{$\sigma_{4}$~(Mb)}&
    \colhead{$\sigma_{5}$~(Mb)}&
    \colhead{$\sigma_{\Sigma}$~(Mb)} &
    \colhead{$\bar q$}
    }
\startdata
\phantom{1}691.568 & 0.1191(87) & 0.0845(59) & 0.0106(24) & 0.0004(04) & -            &   0.214(11) & 4.498(30) \\
\phantom{1}711.001 & 4.976(53)  & 5.557(47)  & 0.542(11)  & 0.0333(30) &  -           &  11.109(71) & 4.6069(36) \\
\phantom{1}711.602 & 4.021(34)  & 4.938(34)  & 0.4561(85) & 0.0264(19) & 0.000492(94) & 9.441(48) & 4.6280(29) \\
\phantom{1}718.213 & 0.573(18)  & 0.921(19)  & 0.0991(47) & 0.0065(13) & -            & 1.600(27) & 4.7119(97) \\
\phantom{1}723.021 & 1.756(31)  & 2.609(32)  & 0.2852(81) & 0.0198(23) & -            & 4.670(45) & 4.6935(57) \\
\phantom{1}731.636 & 0.2067(80) & 0.1873(66) & 0.0204(18) & 0.00195(53)& 0.000149(59) & 0.416(11) & 4.562(15) \\
\phantom{1}744.057 & 0.397(16)  & 1.803(27)  & 0.3603(89) & 0.0222(24) & -            & 2.582(33) & 5.0030(71) \\
\phantom{1}756.678 & 0.384(15)  & 1.509(17)  & 0.460(19)  & 0.0303(28) & 0.000414(93) & 2.384(25) & 5.0574(81) \\
\phantom{1}771.703 & 0.1947(77) & 0.726(13)  & 0.6316(97) & 0.0281(19) & 0.00083(12)  & 1.581(18) & 5.3119(83) \\
\phantom{1}801.754 & 0.1333(64) & 0.637(12)  & 0.789(11)  & 0.0379(21) & 0.00131(12)  & 1.598(18) & 5.4579(80) \\
\phantom{1}901.923 & 0.0986(56) & 0.490(10)  & 0.814(12)  & 0.1036(35) & 0.00387(22)  & 1.506(17) & 5.6125(86) \\
\enddata
\tablenotetext{}{(This table is available in its entirety in machine-readable form.)}
\end{deluxetable*}

The measured partial cross sections, $\sigma_m$, for one- to five-fold ionization of Fe$^{3+}$ are shown in Figure~\ref{fig:cross_sections} and are also presented numerically in Table~\ref{tab:Xsec}. 
They span about six orders of magnitude, ranging from almost $10\,\mathrm{Mb}$ to less than $0.1\,\mathrm{kb}$. 
All measured partial cross sections exhibit a complex resonance structure below the $2p$ ionization threshold. These resonances arise primarily from $2p \rightarrow nd$  excitations located below and slightly above the $2p$ ionization threshold. 
According to our calculation, this threshold is located at 762~eV. 
\citet{Verner1993a} obtained a slightly different value of 766.9~eV. 
We expect our result to be more accurate with an expected uncertainty of only a few eV.
Due to the presence of metastable species in the Fe$^{3+}$ ion beam, the threshold can be expected to be somewhat washed out. 
The 224 fine-structure levels of the $2s^2 2p^5 3s^2 3p^6 3d^5$ configuration of Fe$^{4+}$ span an energy range of about 35~eV from approximately 762 to 797~eV. 
In Figure~\ref{fig:cross_sections} these are represented by vertical gray bars.

The calculations show that the measured resonance structures are often blends of many resonance transitions from the ground level, and from the metastable levels of the ground configuration, to the different $2s^2 2p^5 3s^2 3p^6 3d^5 nd$ core-hole excited levels. The most prominent feature, which can be discerned in the experimental data, is the $2p_{3/2}-2p_{1/2}$ fine-structure splitting of about 15~eV that shows up in the two strong peaks between 700 and 730~eV, where the stronger peak at about 711~eV belongs to excitations of $2p_{3/2}$ electrons.

The resonance structure associated with the $2s \rightarrow np$ ($n \geq 4$) transitions around 870~eV can be seen in all of the ionization channels. They are much weaker than the features associated with $2p$ excitations, as the photoabsorption probability is lower due to the fewer number of electrons in the $2s$ shell. Furthermore, the decay widths of $2s$ core excited states are about a factor 9 larger than the widths of $2p$ holes, due to the rapid Coster--Kronig process where the $2s$ hole is filled by a $2p$ electron. As a consequence, all $2s$ resonances have a much larger width and hence appear much weaker compared to the direct ionization background. 
The $2s$ ionization threshold is expected at 902~eV according to our calculations and at 885~eV according to the work of \citet{Verner1993a}. 
Again, we expect our result to be more accurate, with an uncertainty of only a few eV.
This threshold cannot be directly seen in the experimental data. All 74 fine-structure levels of the $2s 2p^6 3s^2 3p^6 3d^5$  configuration as calculated are shown as gray vertical bars in Figure~\ref{fig:cross_sections}.

The experimental total photoabsorption cross section given by Equation~\eqref{eq:absorption_cross_section} is shown in Figure~\ref{fig:absorption_cross_section} and compared to the photoionization cross section computed by \citet{Verner1993a}. The latter includes only direct single-electron photoionization and therefore the resonance features are absent in the computed cross section.  At energies above the $2p$ and $2s$ resonances the experimental cross section decreases less steeply than the theoretical result. A similar behavior was also observed for Fe$^+$ \citep{Schippers2017}, albeit over a much narrower energy range. Here the deviation between the experimental photoabsorption cross section and the result of \citet{Verner1993a} reaches almost a factor of $\mathord{\sim} 1.5$ at the highest experimental photon energy of 950~eV. At present, the reason for this discrepancy is not known. One might speculate that the population of metastable levels in the primary ions leads to a change of the photoionization cross section. However, a strong change of the inner-shell ionization cross section upon excitation of the outermost electrons by only a few eV does not seem very likely.

\begin{figure}
\centering
\includegraphics[width=1.0\linewidth]{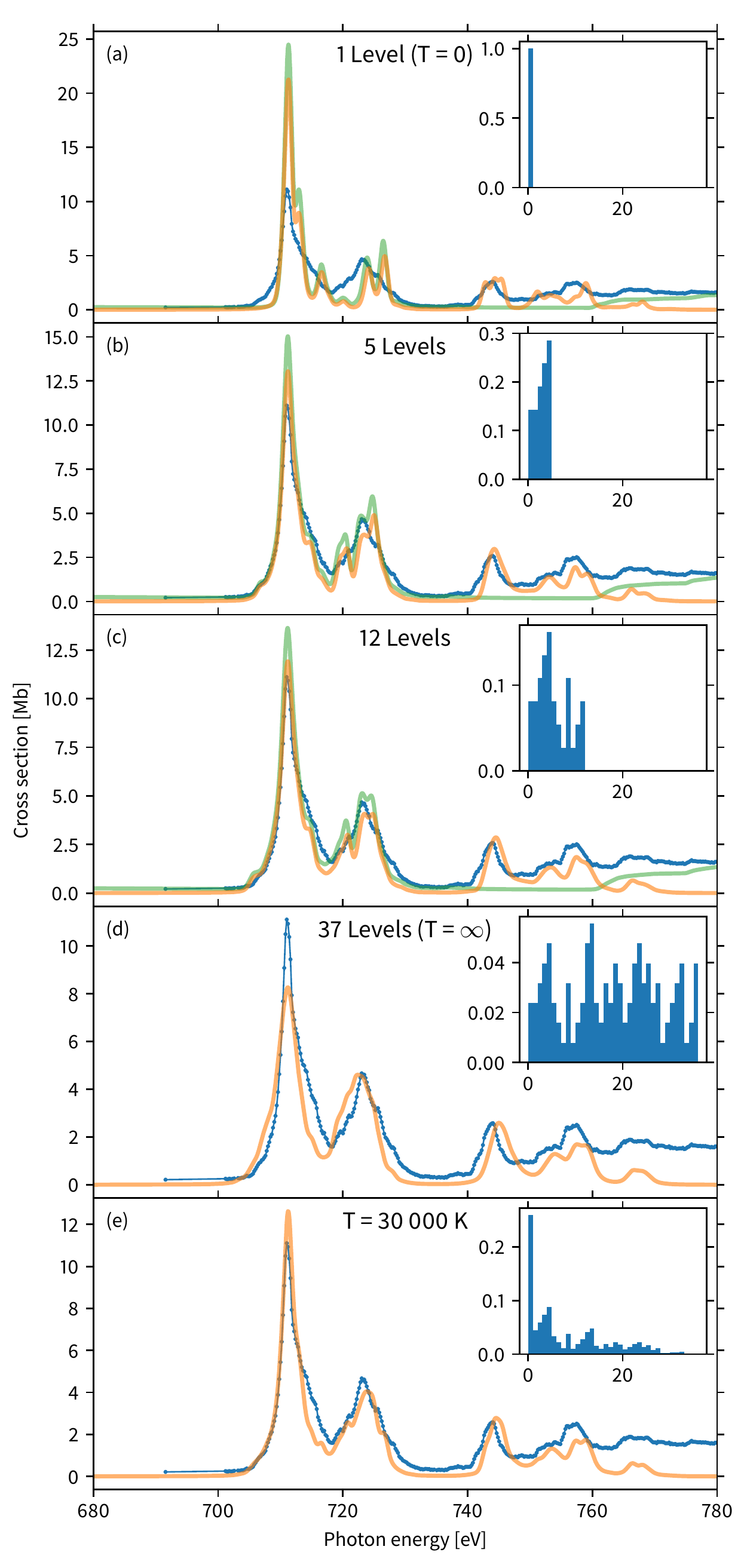}
\caption{Computed cross section for different populations of the $3d^5$ ground configuration.
Panels (a) -- (d) are computed with a statistical population of the lowest $N = 1,5,12,$ and $37$ fine-structure levels, respectively, while (e) is based on a Boltzmann distribution at $T = 30\,000 \, \mathrm{K}$.
The blue dots are the experimental data from Figure~\ref{fig:absorption_cross_section}, the orange line represents the MCDHF computation of the cross section for photoexcitation and the green curves are the HFR photoabsorption cross sections (photoionization and photoexcitation) not including $4d$ and $5d$ excitations.
The computed spectra were convoluted with a Voigt profile with a Gaussian FWHM of 1.0~eV and lifetime broadening of $\Gamma = 0.4$~eV.
The computed MCDHF and HFR energies have been shifted by $-2.2$~eV.
The inset histograms show the relative population vs.\ level number for the 37 ground-configuration fine-structure levels that is assumed for each plot.}
\label{fig:absorption_population}
\end{figure}

\subsection{Photoabsorption Cross Section}

Using our calculations, we investigated the effects on our theoretical cross sections due to different populations of the 37 levels of the ground configuration. 
In each panel of Figure~\ref{fig:absorption_population}, we compare the experimental photoabsorption cross section, shown in blue, with MCDHF and HFR results based on different populations of the fine-structure levels in the ground configuration. 
These MCDHF results include only photoexcitations into the $3d, 4d, 5d$ shells. 
The HFR results omit contributions from the $4d$ and $5d$ shells but also include photoionization of $M$ and $L$-shell electrons.
The increase of the HFR cross section starting around 760~eV is the contribution from the photoionization of $2p$ electrons.
The respective level populations are displayed in the insets of the panels.
In order to account for the uncertainty due to the experimental photon-energy spread and the lifetime broadening, the computed data were convoluted with a Voigt profile, where the full width at half maximum (FWHM) of the Gaussian was chosen as 1.0~eV and a uniform natural line width of $\Gamma = 0.4$~eV was assumed. 
In addition, the calculated spectra were shifted by $-2.2$~eV  such that the theoretical and experimental positions of the tallest resonance feature at about 711~eV match.

\begin{figure}[t]
\includegraphics[width=\linewidth]{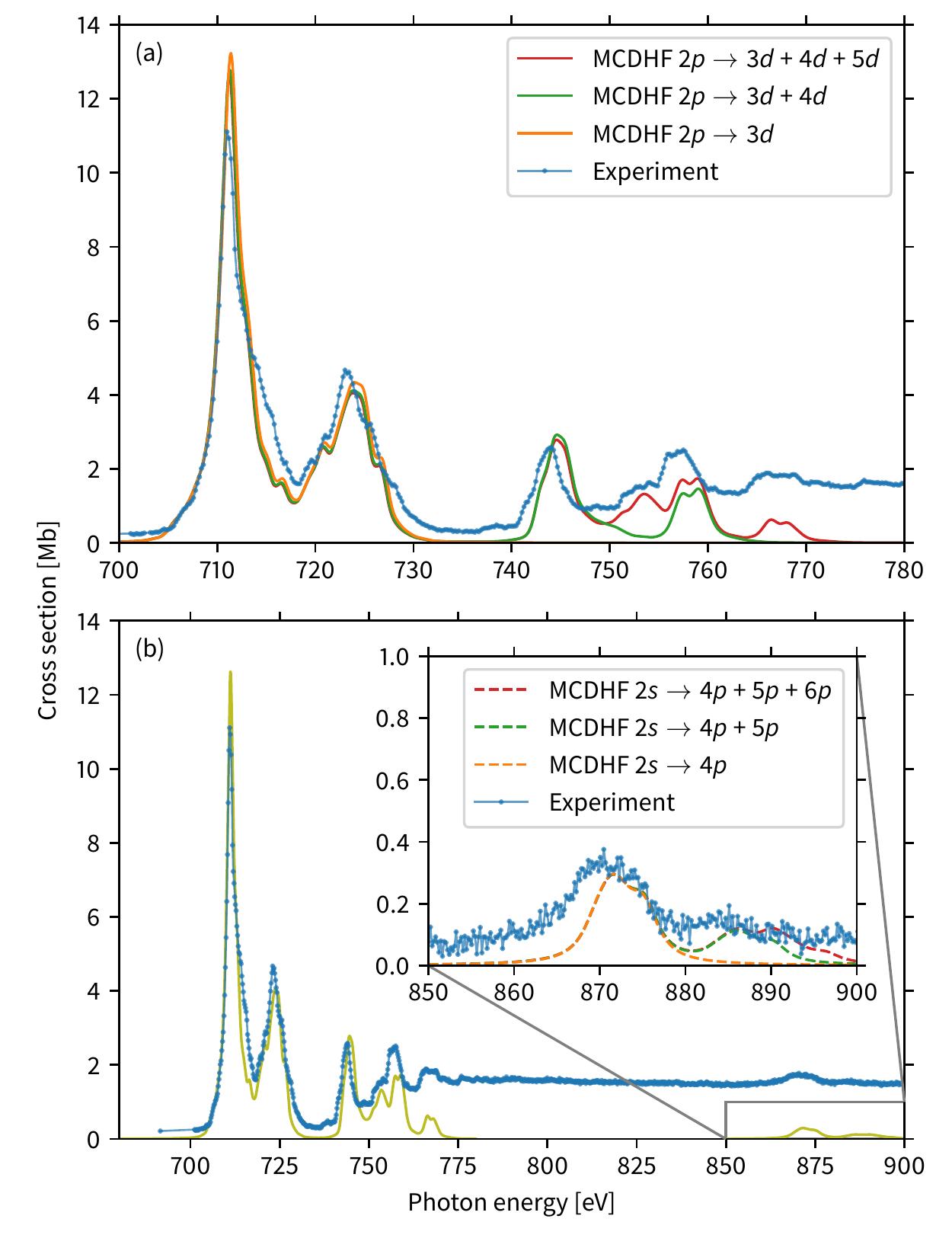}
\caption{Measured photoabsorption cross section and our MCDHF calculations, including (a) $2p \to nd$ resonances and (b) $2s \to np$ resonances. 
The inset in (b) enlarges the region of the $2s\to np$ resonances. 
Computed spectra are convoluted with a Voigt profile with a Gaussian $\mathrm{FWHM} = 1.0$~eV and natural line widths of $\Gamma = 0.4$~eV and $\Gamma = 3.5$~eV for the $nd$ and $np$ resonances, respectively. 
Here, an offset of $-1.4$~Mb was added to the experimental data in order to reduce the $L$ and $M$-shell photoionization background and to facilitate a comparison with the computation.
The computed MCDHF energies have been shifted by $-2.2$~eV. 
Where the theoretical curves overlap, the orange curve lies on top of the green curve, which lies on top of the red curve.
}
\label{fig:absorption_cross_section_detail}
\end{figure}

In the top panel (Figure~\ref{fig:absorption_population}a), we assume that only the well separated ground level is populated in the initial ion beam. As a consequence, both the MCDHF and HFR calculations overpredict the cross section, especially for the $2p_{3/2}$ excitation at about 711~eV. Moreover, the calculated cross sections exhibit more details than the experimental photoabsorption spectrum. Both theories agree very well with each other. However, the $4d$ and $5d$ excitations were not included in the HFR calculations and therefore the corresponding resonances are only visible in the MCDHF results. Furthermore, CI between the different $nd$ configurations slightly reduces the MCDHF cross section, as can also be seen in Figure~\ref{fig:absorption_cross_section_detail}. This partially accounts for the lower peak cross section predicted by our MCDHF results as compared to the HFR results seen in Figure~\ref{fig:absorption_population}.

In Figure~\ref{fig:absorption_population}b we assume the statistical population of the $3d^5 \, {}^6S_{5/2}$ ground level and the $3d^5 \, {}^4G$ first excited multiplet, as seen in the inset. As a consequence, both theories predict that some of the fine structure that is visible in Figure~\ref{fig:absorption_population}a cannot be resolved anymore and that the strongest line becomes wider, while its maximum is drastically lowered, in better agreement with the experiment. The same trend continues, when the next two multiplets (${}^4P$ and ${}^4D$) are included in the statistical mixture, as seen in Figure~\ref{fig:absorption_population}c. Compared to the experimental results, the total theoretical cross sections are in good agreement, though too much fine structure still remains visible in the theory. When the statistical average is extended over all 37 fine-structure levels of the ground configuration, the remaining fine structure also vanishes and only 6 rather broad lines remain, as seen in  Figure~\ref{fig:absorption_population}d.
Also noteworthy is that the $2p_{3/2}$ resonance feature is underestimated in this model.

These results show that the assumption of just the ground level being populated is not justified, neither is the assumption of a statistical population of all levels in the ground configuration. Furthermore, a drastic cut in the population, such as in Figures~\ref{fig:absorption_population}b and \ref{fig:absorption_population}c is also a rather unrealistic scenario, especially since only the ground level is energetically well separated. Therefore, a population that gives clear preference to the ground level but also populates all excited levels of the ground configuration seems more appropriate. For this purpose we chose a Boltzmann distribution at a temperature of 30\,000~K with no other justification than the relatively good agreement between the calculated and measured photoabsorption spectra, as seen in  Figure~\ref{fig:absorption_population}e. 
This temperature also seems plausible in view of the electron energies that have been estimated for plasmas in ECR ion sources \citep{Trassl2003}.
At this temperature, the population within any given multiplet is almost statistical, while the population of excited multiplets is suppressed due to their high excitation energies. The result of choosing this distribution and temperature is in good agreement with the experimental results, not only in terms of the maximum value of the cross section but also for the width of the resulting lines. All the following results were computed with this distribution for the population of the 37 fine-structure levels of the ground configuration in the ion beam.

The positions of the photoexcitation peaks also slightly depend on the population of metastable levels in  the ion beam. The strongest shift is observed for the line around 722~eV which is a blend of many transitions. Here, the shift between Figures~\ref{fig:absorption_population}d and \ref{fig:absorption_population}e is about 1.4~eV. For the line at approximately 745~eV, which primarily arises from $2p_{3/2} \rightarrow 4d$ excitations, the shift is about 0.5~eV. The position of the tallest peak at 711~eV, which is associated with $2p_{3/2} \rightarrow 3d$ excitations, however, is almost constant, shifting by 0.1~eV at most.

Figure~\ref{fig:absorption_cross_section_detail}a shows the experimental photoabsorption cross section in the $2p$-threshold region together with the computed photoexcitation cross section resulting from $2p \rightarrow nd$ ($n = 3,4,5$) excitations. 
The three lowest lines arise from $2p \rightarrow 3d$ or $4d$ excitations while the higher resonance structures are blends of contributions with different principal quantum numbers of the upper levels. 

Figure~\ref{fig:absorption_cross_section_detail}b displays the measured and computed cross sections over a larger energy range, that also includes the $2s$ threshold. The cross sections around the $2p$ threshold are identical to the ones in Figure~\ref{fig:absorption_cross_section_detail}a, while the computed data for the $2s$ core excited levels (inset) are convoluted with a Voigt profile with a Lorentzian width $\Gamma = 3.5$~eV in order to account for the much faster decay of those states (cf., Sec.~\ref{sec:mcdf}). Again, the lowest three $np$ ($n = 4,5,6$) shells were taken into account. As seen from the inset of this figure, these contributions are also visible in the experimental data.

\subsection{Product Charge State Fractions}\label{sec:fractions}

\begin{figure*}[t]
\includegraphics[width=\linewidth]{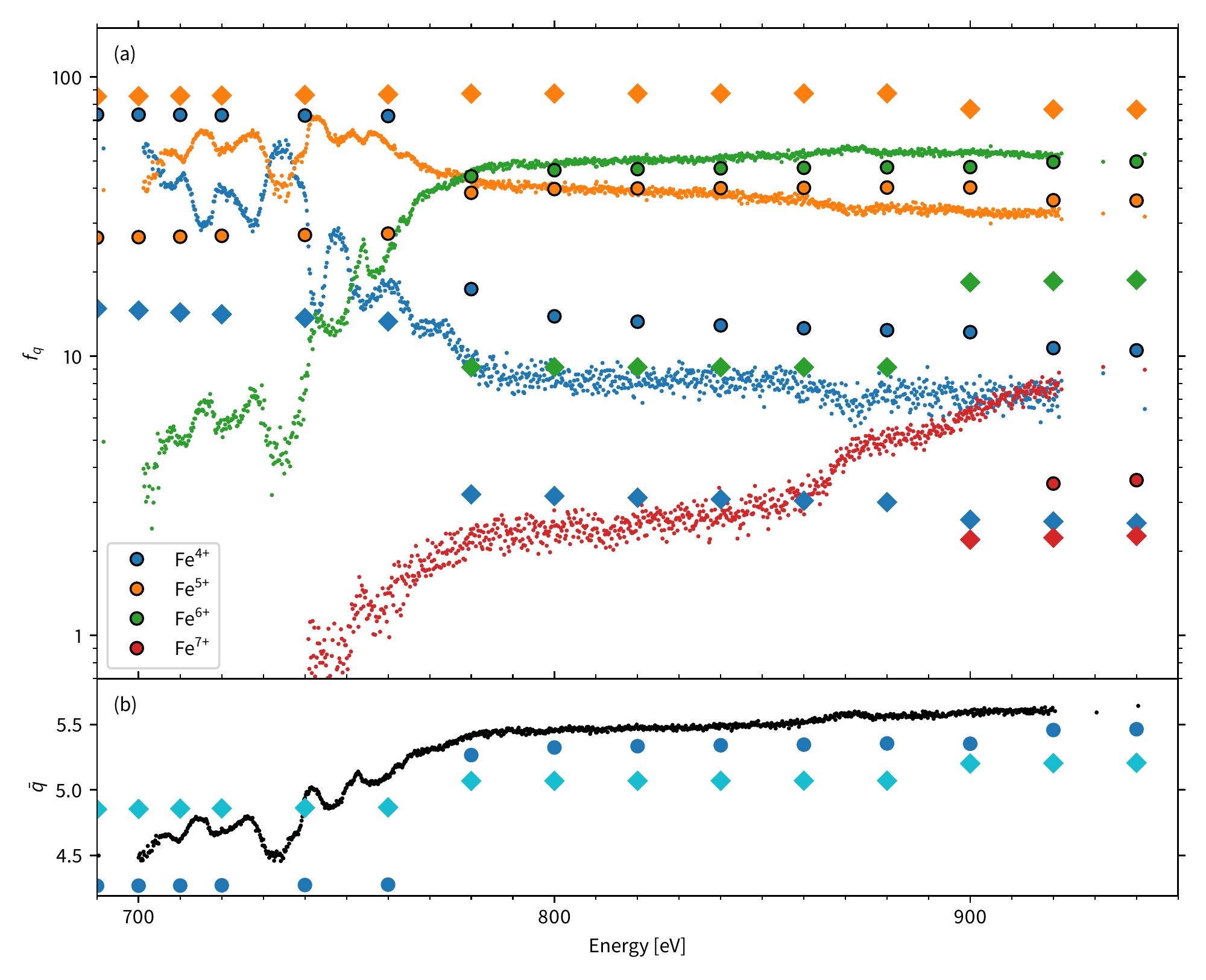}
\caption{\label{fig:total_ion_yield}(a) Product charge-state fractions, $f_q$, in percentage for the four charge states $q = 4,5,6$, and $7$. Experimental results (small circles) are compared to our computations (large circles) for direct ionization of a single electron using the shake-down Auger model and to the results by \citet{Kaastra1993} weighted by the relative direct photoionization cross sections of \citet{Verner1993a} (diamonds). (b) Mean charge state from the experimental data (small black circles) and our cascade calculations (large blue circles). The diamonds are again the results from \citet{Kaastra1993} combined with the cross sections of \citet{Verner1993a}.}
\end{figure*}

The product charge-state fractions, i.e., the probabilities of an atom to decay into charge state $q$, can be derived as $f_q\left(E_{\mathrm{ph}}\right) = \sigma_q/\sigma_\Sigma$. 
Here $\sigma_q$ are the measured partial cross sections and $E_{\mathrm{ph}}$ signifies the photon energy.
The key feature of the $f_q$ values is that the systematic uncertainty of the absolute cross section scale cancels out. Furthermore, the $f_q$ fractions can be used to calculate the mean product charge state as

\begin{equation}\label{eq:qmean}
\bar q\left( E_{\mathrm{ph}} \right) = \sum_{q = 4}^8 q f_q = \frac{1}{\sigma_\Sigma} \sum_{m = 1}^5 \left( m + 3 \right) \sigma_m.
\end{equation}

Figure~\ref{fig:total_ion_yield}a shows the product charge-state fractions for the overall ionization process and Figure~\ref{fig:total_ion_yield}b the mean charge state $\bar q$ (see also Table~\ref{tab:Xsec}). In addition to the experimental data, which are displayed by small circles, both figures compare our computed results for these quantities (large circles) with the results obtained as a combination of the theoretical cross sections for photoionization by \citet{Verner1993a} and the cascade calculations by \citet{Kaastra1993} (diamonds).

Here we compute the theoretical product charge-state fractions due either to photoionization or photoexcitation.  
Because of the above mentioned issues in the computation of absolute photoionization cross sections, we did not add together the contributions from photoionization and photoexcitation.  

When considering only photoionization, we calculate the product charge-state fractions using

\begin{equation}\label{eq:charge_state_fractions}
f_q \left(E_{\mathrm{ph}}\right) = \frac{1}{\sigma_{\mathrm{tot}} \left( E_{\mathrm{ph}} \right) } \sum_k \sigma_k \left( E_{\mathrm{ph}} \right) F_{k, q} \, ,
\end{equation}
where $\sigma_k \left( E_{\mathrm{ph}} \right)$ is the cross section for direct photoionization of an electron from subshell $k$ vs.\ photon energy, and the total photoionization cross section is again obtained by summing over all subshells $\sigma_{\mathrm{tot}}\left( E_{\mathrm{ph}} \right) = \sum_k \sigma_k\left( E_{\mathrm{ph}} \right) $. 
$F_{k, q}$ denotes the fraction Fe$^{q+}$ produced after the removal of an electron from subshell $k$ of Fe$^{3+}$ and is discussed in the next subsection.

\begin{table}
\centering
\caption{Comparison of the theoretical photoionization branching ratios $\sigma_k/\sigma_\mathrm{tot}$ from this work with the results of \citet{Verner1993a}.
The results are given in percentage.}
\label{tab:photo_branchings}
\begin{tabular}{cDDDDD}
\midrule \midrule
Energy [eV] & \multicolumn{2}{c}{$2s$} & \multicolumn{2}{c}{$2p$} & \multicolumn{2}{c}{$3s$} & \multicolumn{2}{c}{$3p$} & \multicolumn{2}{c}{$3d$}
\\
\midrule
\decimals
\multicolumn{11}{c}{This work}
\\
\midrule
 690 &  0 &  0 & 20   &  68 & 13 \\
 840 &  0 & 82 &  4   &  12 & 1.8 \\
 960 & 12 & 74 &  3.5 &  10 & 1.2 \\
\midrule
\multicolumn{11}{c}{\citet{Verner1993a}}
\\
\midrule
 690  &  0   &  0   & 20    & 66    & 15 \\
 840  &  0   & 87   &  2.7  &  8.4  & 1.5 \\
 960  & 13   & 76   &  2.5  &  7.4  & 1.1 \\
\midrule
\end{tabular}
\end{table}

The quantities $\sigma_k \left( E_{\mathrm{ph}} \right) / \sigma_{\mathrm{tot}}\left( E_{\mathrm{ph}} \right)$ represent the photoionization branching ratios. We utilized the \textsc{Photo} component of the \textsc{Ratip} code \citep{Fritzsche2012a} to compute these quantities from our MCDHF wave functions for all subshells for which ionization is possible in the given energy range.
In the upper part of Table~\ref{tab:photo_branchings}, we show these results for three energies that are representative for the three main regions covered in the experiment: below the $2p$ threshold, between the $2p$ and $2s$ threshold, and above the latter. 
At these energies, photoionization dominates over photoexcitation.
The lower part of Table~\ref{tab:photo_branchings} shows the theoretical results obtained by \citet{Verner1993a}, using a relativistic Hartree--Dirac--Slater method. Generally, their findings agree well with our results. The rather small differences could be due to differences in the treatment of relaxation effects.

When considering only photoexcitation, we replace $\sigma_k \left( E_{\mathrm{ph}} \right) / \sigma_{\mathrm{tot}}\left( E_{\mathrm{ph}} \right)$ in Equation~\eqref{eq:charge_state_fractions} with the theoretical fractional populations from the photoexcitation transition rates.
The definition of $F_{k, q}$ remains unchanged.

\begin{table}
\centering
\caption{Computed branching fractions $F_{k, q}$, given here in percentage, of an inner shell hole created in subshell $k$ by direct (single) photoionization.}
\label{tab:hole_branchings}
\begin{tabular}{lDDDDD}
\midrule \midrule
$k/q$ & \multicolumn{2}{c}{Fe$^{4+}$} & \multicolumn{2}{c}{Fe$^{5+}$} & \multicolumn{2}{c}{Fe$^{6+}$} & \multicolumn{2}{c}{Fe$^{7+}$} & \multicolumn{2}{c}{Fe$^{8+}$}\\
\midrule
\decimals
\multicolumn{11}{c}{This work (shake-down)} \\
 $2s$ &   .  &  2.5 & 64   & 33.6 &  . \\
 $2p$ &   .  & 47   & 53   &  .   &  . \\
 $3s$ &   .  & 100  &  .   &  .   &  . \\
 $3p$ & 100 &   .   &  .   &  .   &  . \\
\midrule
\multicolumn{11}{c}{This work (two-electron Auger)} \\
$2s$ &  .   &  4.0 & 95    & 1.1   & . \\
$2p$ &  .   & 89   & 11    &   .   & . \\
$3s$ &  .   & 100  &   .   &   .   & . \\
$3p$ & 100 &   .   &   .   &   .   & . \\
\midrule
\multicolumn{11}{c}{\citet{Kaastra1993}} \\
$2s$        &  .   &   0.3 & 83.0  & 14.3  & 0.04 \\
$2p_{1/2}$  &  1.8 &  87.2 & 10.5  &  0.54 &  .   \\
$2p_{3/2}$  &  1.1 &  84.9 & 13.3  &  0.67 &  .   \\
$3s$        &   .  & 100   &   .   &  .    &  .   \\
$3p$        &   .  & 100   &   .   &  .    &  .   \\
\midrule
\end{tabular}
\end{table}

\subsection{Cascade Models} \label{sec:cascades}

The branching fractions $F_{k, q}$ were computed for all inner-shell holes that can be created at the photon energies under consideration by utilizing the MCDHF cascade calculations explained in Section \ref{sec:mcdf}.
Previous cascade calculations were performed by \citet{Kaastra1993} to predict the branching fractions after inner-shell ionization for various transition metal elements.
Their results for Fe$^{3+}$  are shown in the lowest part of Table~\ref{tab:hole_branchings}.

In our most straight-forward Auger model, we built the cascade tree by including all energetically allowed two-electron Auger processes. 
The results from this model, denoted as two-electron Auger, are shown in the middle part of Table~\ref{tab:hole_branchings}. 
They agree to a large extent with the earlier results of \citet{Kaastra1993}. One notable exception concerns the decay of $3p$ holes. According to our computations, the corresponding high-lying levels are above the ionization threshold (cf., Figure~\ref{fig:level_Diagram}), but they do not get populated to a significant extent in the photoionization process, so that almost all $3p$ holes formed produce only Fe$^{4+}$. In contrast, \citet{Kaastra1993} find that a $3p$ hole will autoionize and, thus, lead to the formation of Fe$^{5+}$.

For the higher product charge states, there are several inner-shell hole configurations that, for energetic reasons, are partially forbidden to decay via two-electron Auger processes. 
Figure~\ref{fig:level_Diagram} displays three examples that are marked in red and that arise in the decay of the $2s^2 2p^5 3s^2 3p^6 3d^6$ configuration (configuration 2 in Figure~\ref{fig:level_Diagram}). 
For example, the higher-lying fine-structure levels of the $2s^2 2p^6 3s^2 3p^4 3d^6$ configuration (configuration 7 in Figure~\ref{fig:level_Diagram}) can decay via a two-electron Auger process to $2s^2 2p^6 3s^2 3p^5 3d^4$ (configuration 13 in Figure~\ref{fig:level_Diagram}), while this decay path is forbidden for the lower lying levels.
However, these lower levels are still above the ionization threshold for Fe$^{4+}$ forming Fe$^{5+}$. 
Therefore, they can decay by a three-electron Auger process where a third electron undergoes a shake-down $3d \rightarrow 3p$ transition filling the $3p^4$ double vacancy and thereby forming the ground configuration of Fe$^{5+}$ (configuration 12 in Figure~\ref{fig:level_Diagram}). 
In general, such three-electron Auger processes are expected to be slow compared to a two-electron Auger process. Nevertheless, they can still be faster than the competing radiative processes that would result in Fe$^{4+}$ product ions. The precise computation of the Auger transition rates including a shake-down transition is rather challenging due to complex correlation patterns \citep{Andersson2015, Schippers2016a, Beerwerth2017}. Here we assume that the radiative losses are still negligible, so that all levels that are energetically allowed to autoionize will do so.
In the following we will refer to this extended cascade decay tree as \lq\lq{}shake-down\rq\rq. The resulting branching fractions $F_{k, q}$ are shown in the upper part of Table~\ref{tab:hole_branchings}. They give rise to drastic changes in the ion yield from  $2p$ and $2s$ holes.
For example, the yields of Fe$^{6+}$ and Fe$^{7+}$, respectively, are significantly increased.

\begin{table}
\centering
\caption{Experimental and theoretical product charge-state fractions $f_q$ upon photoexcitation or direct photoionization of Fe$^{3+}$ by a photon of the given energy.
The results are given in percentage.}
\label{tab:ion_yield}
\begin{tabular}{lDDDDD}
\midrule  \midrule
{Energy [eV]} &  \multicolumn{2}{c}{Fe$^{4+}$} &  \multicolumn{2}{c}{Fe$^{5+}$} &  \multicolumn{2}{c}{Fe$^{6+}$} &  \multicolumn{2}{c}{Fe$^{7+}$} & \multicolumn{2}{c}{Fe$^{8+}$} \\
\midrule
\decimals
\multicolumn{11}{c}{$2p \rightarrow 3d$ resonances (experiment)} \\
711   & 45. & 50. & 5. & 0.3 & 0.01 \\
723   & 34. & 59. & 7. & 0.3 & 0.01 \\
\midrule
\multicolumn{11}{c}{$2p \rightarrow 3d$ resonances (shake-down)} \\
711  & 41.9 & 56.7 & 1.3 & . \\
723  & 40.3 & 56.9 & 2.7 & . \\
\midrule
\multicolumn{11}{c}{$2p \rightarrow 3d$ resonances (two-electron Auger)} \\
711  & 66.9 & 31.7 & 1.3 & . \\
723  & 54.6 & 42.7 & 2.8 & . \\
\midrule
\multicolumn{11}{c}{Direct ionization (experiment)} \\
  690 & 55.6 & 39.4 &  4.9 & 0.2 & 0.001 \\
  840 &  7.7 & 38.2 & 51.4 & 2.7 & 0.1 \\
  960 &  6.9 & 32.2 & 51.5 & 9.4 & 0.4 \\
\midrule
\multicolumn{11}{c}{Direct ionization (shake-down)} \\
  690 & 73.4 & 26.6 & 0.0  & 0.0 \\
  840 & 12.9 & 40.0 & 47.1 & 0.0 \\
  960 & 10.4 & 35.9 & 49.9 & 3.8 \\
\midrule
\multicolumn{11}{c}{Direct ionization (two-electron Auger)} \\
  690 & 73.4 & 26.6 &  0.0 & 0.0 \\
  840 & 12.9 & 76.9 & 10.1 & 0.0 \\
  960 & 10.4 & 69.2 & 20.2 & 0.2 \\
\midrule
\end{tabular}
\end{table}

We can combine the fractions $F_{k, q}$ with the computed photoionization branching ratios $\sigma_k/\sigma_\mathrm{tot}$ from Table~\ref{tab:photo_branchings} in order to model the full decay tree and compare the resulting ion yields and mean charge state vs.\ photon energy to the experimental results. 
The resulting product charge-state fractions are given in Table~\ref{tab:ion_yield} for both photoexcitation of the initial ion as well as for direct photoionization. 
For both cases, the results are again given for the two cascade models introduced before, with and without shake-down transitions included. 
In the case of direct ionization, the results are given for three energies, below the $2p$ threshold, between the $2p$ and $2s$ thresholds, and above the latter. 
As already expected from the ion fractions $F_{k, q}$ in Table~\ref{tab:hole_branchings}, the total product charge-state fractions from the two models differ dramatically.

The theoretical product charge-state fractions due to photoionization only are graphically presented in Figure~\ref{fig:total_ion_yield}, together with the experimental data. 
The small circles are the experimental data, while the large circles are our theoretical values using the shake-down Auger model. 
Our theoretical data do not reproduce the measured resonance structures because we account only for photoionization here and do not include the effects of photoexcitation.
The diamonds are the theoretical results that are obtained by combining the photoionization branchings from \citet{Verner1993a} with the cascade calculations by \citet{Kaastra1993}. 
For this last case, the resulting charge-state fractions disagree significantly with the experiment.
This was also seen for the respective calculations for Fe$^+$ by \citet{Schippers2017}. 
The mean charge state from the combined \citet{Verner1993a} and \citet{Kaastra1993} results is significantly overestimated below the $2p$ ionization threshold and the step at the ionization threshold is much less pronounced than in the experimental data. 
Above the $2p$ ionization threshold, the mean charge state is significantly underestimated. This behavior arises because the calculations by \citet{Kaastra1993} predict the fraction of Fe$^{5+}$ to be about a factor of two too high, while the predicted fraction of Fe$^{6+}$ is about an order of magnitude too low.
Similarly, both the predicted Fe$^{4+}$ and Fe$^{7+}$ charge-state fractions are also too low. 
The low Fe$^{4+}$ fraction is a consequence of the autoionizing behavior of $3p$ holes that was predicted by \citet{Kaastra1993} and that disagrees with our present findings.

Figure~\ref{fig:total_ion_yield} shows that our calculations represent a significant improvement over the previous computations by \citet{Kaastra1993}.
Most notably, as can be seen in Figure~\ref{fig:total_ion_yield}b, the pronounced step in the mean charge state at the $2p$ ionization threshold is clearly reproduced and is hence in much better agreement with experiment, but still somewhat underestimated.
As can be seen in Figure~\ref{fig:total_ion_yield}a, our calculations also predict the charge-state fractions more accurately than the previous theory.
Most importantly, the two strongest channels, Fe$^{6+}$ and Fe$^{5+}$, are predicted quite well and in the correct order.
However, the production of Fe$^{4+}$ is still slightly overestimated, and the production of the highest measured charge states ($q = 7, 8$) is significantly underestimated.
The main reason that our computations are in better agreement with the experiment than previous theory is the incorporation of shake-down transitions and of more precise transition energies and rates from our fine-structure resolved treatment.

\section{Summary and Conclusions}\label{sec:summary}

We have measured relative cross sections for up to five-fold ionization of Fe$^{3+}$ ions after resonant $L$-shell photoexcitation or direct photoionization.
We have used a photon-ion merged-beams technique. The present measurements are a continuation of the earlier work on Fe$^{+}$ \citep{Schippers2017}. 
We observed strong ionization resonances due to $2p \rightarrow nd$ excitations, where contributions by $n = 3,4$, and $5$ could be identified with the help of MCDHF calculations. Around the $2s$ ionization threshold, we were able to identify $2s \rightarrow np$ resonances, where the $4p$ contribution can be clearly seen and higher shells contribute to some weak and broad feature.

Furthermore, we performed extensive calculations of the de-excitation cascades that follow upon the creation of holes in the $2s$ and $2p$ shells. Our computed product charge-state fractions agree well with the experimental results, where we found that the contribution of several three-electron Auger processes is the likely main reason why earlier theory based on cascade branching fractions by \citet{Kaastra1993} and photoionization cross sections by \citet{Verner1993a} fail to reproduce the current experimental results. 
Despite these improvements, our current Auger models show notable deficiencies in describing the formation of the highest charge states, in this case Fe$^{7+}$ and Fe$^{8+}$.
In our models Fe$^{8+}$ is not included due to energy conservation, and important decay paths leading to Fe$^{7+}$ are still missing.
The starting point for including these charge states into an Auger model would be to include shake-up transitions in the Auger decay of $2p$ vacancies or direct double Auger decay processes of $2p$ vacancies.

The computation of the photoabsorption spectra is complicated by the presence of ions in metastable levels in the experiment. From the comparison with experiment, this effect is found to be more severe than in the previous study on Fe$^{+}$.
Still even with these experimental issues, computations of resonant photoabsorption spectra agree reasonably well with experiment when all 37 fine-structure levels of the ground configuration are assumed to be populated at a temperature of 30\,000~K in the ion beam. Additionally, since the Fe$^{3+}$ resonance positions are significantly different from the Fe$^+$ resonance positions published before \citep{Schippers2017}, it should still be possible to identify individual signatures from both charge states in X-ray photoabsorption or emission spectra. We will discuss this aspect in more depth in a future publication where we will also present experimental and theoretical data for single and multiple ionization of Fe$^{2+}$.
Lastly, our benchmarked theoretical results are also being incorporated into models for X-ray absorption in the ISM (T. Kallman, private communication) and are available upon request.

\newpage

\acknowledgments

This research was carried out at the light source PETRA\,III at DESY, a member of the Helmholtz Association (HGF).
We would like to thank G.~Hartmann, F.~Scholz, and J.~Seltmann, for assistance in using beamline P04.
This research has been funded in part by the German Federal Ministry for Education and Research (BMBF) within the \lq\lq{}Verbundforschung\rq\rq\ funding scheme under contracts 05K16GUC, 05K16RG1, and 05K16SJA.
A.M. acknowledges support by Deutsche Forschungsgemeinschaft. 
S.B. and K.S. would like to thank SFB 755, \lq\lq{}Nanoscale photonic imaging, project B03\rq\rq\ for financial support and acknowledge funding from the Initiative and Networking Fund of the Helmholtz Association.
D.W.S. was supported, in part, by the NASA Astrophysics Research and Analysis program and the Astrophysics Data Analysis Program.


\bibliographystyle{aasjournal}

\end{document}